*Ab initio* prediction of Boron compounds arising from Borozene: Structural and electronic properties

G. Forte[a,*], A. La Magna[b], I. Deretzis[b,c] and R. Pucci[d]

[a] *Dipartimento di Scienze Chimiche, Facoltà di Farmacia, Università di Catania, Viale Doria 6, I-95126 Catania, Italy*

[b] *CNR-IMM, I-95121 Catania, Italy*

[c] *Scuola superiore, Università di Catania, I-95123 Catania, Italy*

[d] *Dipartimento di Fisica e Astronomia, Università di Catania, 95123, Catania, Italy*

**Abstract**

Structure and electronic properties of two unusual boron clusters obtained by fusion of borozene rings has been studied by means of first principles calculations, based on the generalized-gradient approximation of the density functional theory, and the semiempirical tight-binding method was used for the transport calculations. The role of disorder has also been considered with single vacancies and substitutional atoms. Results show that the pure boron clusters are topologically planar and characterized by (3c-2e) bonds, which can explain, together with the aromaticity (estimated by means of NICS), the remarkable cohesive energy values obtained. Such feature makes these systems competitive with the most stable boron clusters to date. On the contrary, the introduction of impurities compromises stability and planarity in both cases. The energy gap values indicate that these clusters possess a semiconducting character, while when the larger system is considered, zero-values of the density of states are found exclusively within the HOMO-LUMO gap. Electron transport calculations within the Landauer formalism confirm these indications, showing semiconductor-like low bias differential conductance for these stuctures. Differences and similarities with Carbon clusters are highlighted in the discussion.

[*] Corresponding author. Tel. +39 095 738 5066; email address: gforte@unict.it.

# 1. Introduction

Boron is the first element in group IIIA of the periodic table, presents the external electronic configuration $s^2p^1$ and possesses a variety of compounds second only to carbon.

In several boron compounds [1,2] the existence of multicenter bonds has been discovered, which arise from the electron deficiency of this element. Moreover boron has a special place among the elements of the periodic table because of the wide variety of crystalline structure forms, i.e. polymorphism, which include nanotubes [3], nanoribbons [4] and nanoclusters [5,6].

The interest in boron-based nanostructures has recently increased due to new studies of both the synthesis of single walled boron nanotubes (SWBNTs) and the prediction of ballistic conduction in SWBNTs [3,6]; these findings, together with the properties that all boron nanotubes a) are predicted to be metallic [7] and b) are superconducting at low temperatures [8,9], promoted their prospective applications in fabrication of novel electronic devices.

The most stable boron structure is the α-rhombohedral bulk where boron icosahedra are centered on the edges of a rhombohedral unit cell [10]. Unlike the bulk boron compounds, boron clusters $B_n$ (n < 20) are quasiplanar, or even planar, with a symmetrical bond distribution, aromatic [11-13] and their existence is confirmed by the experiment [14]. From the Aufbau principle postulated by Boustani [5] it follows that these quasiplanar isomers are more stable than their icosahedral counterparts.

Recently Szwacki *et al.* have predicted the existence of a planar and aromatic boron compound, named borozene, which has strong similitudes with benzene [15].

Motivated by these findings we present a work regarding a first principles study, within the generalized gradient approximation (GGA), in terms of structural and electronic properties of two boron compounds, $B_{60}H_{12}$ and $B_{228}H_{24}$, in which the molecule of borozene can be considered as the building block, as the benzene ring represents the embryo of compounds such Coronene, Coronene 19 etc.

In general we will refer to these compounds as boron clusters whose external dangling bonds are saturated by hydrogen atoms; they are obtained by fusing together the outer boron pairs of borozene molecules bonded to a hydrogen atom, see **Fig.1.**

The effect of impurities (carbon or nitrogen substitutional atoms) and vacancies will also be discussed, and differences or similarities with carbon clusters will be emphasized.

We have organized the rest of the paper as follows: the computational methods adopted are presented in Section 2. In Section 3 we present and discuss our results in comparison also with carbon compounds, and, finally, we give a summary in section IV.

## 2. Computational Method.

The molecule $B_{60}H_{12}$ here considered was built by fusing six borozene rings, for this reason it can be considered as the boron counterpart of coronene, whereas the structure of $B_{228}H_{24}$ was obtained by surrounding $B_{60}H_{12}$ with one series of borozene rings, therefore this cluster is constituted by a total of 24 borozene rings.

Carbon and nitrogen were chosen as substitutional impurities since both can easily undergo $sp^2$ hybridization, allowing the aggregation of stable planar structures; a single vacancy was obtained by removing one of the inner boron atoms.

A first optimization energy procedure was performed in the framework of the molecular mechanics approximation applying the CVFF Force Field [16,17] which is enclosed in the Materials Studio package [18]. The geometries obtained were fully optimized at a B3LYP/STO-3G [19-23] and B3LYP/6-311G [24-25] level by using the quadratically convergent Self Consistent Field procedure [26].

In detail, due to the large size, the optimizations of $B_{228}H_{24}$ and its analogues with impurities, were carried out by means of the minimal basis set STO-3G whereas the more extended Pople basis set 6-311G was used in the optimizations of $B_{60}H_{12}$ and related molecules with impurities.

In order to estimate the degree of aromaticity, the calculation of Nuclear Independent Chemical Shifts [27] on the plane of the aromatic system (NICS0) was computed using the Gaussian 03 package [28], applying the GIAO method [29,30]. To obtain the contour plot of NICS, ghost atoms were placed on the plane of the molecule with a step size of about 1A.

Finally, electronic transport has been evaluated in the framework of the Nonequilibrium Green's Function theory using a Landauer expression for the calculation of the current-voltage I-V characteristics [31]. Consistently to the electron structure findings we assume that only $p_z$ orbitals contribute to the low-bias transport along the molecules and that the Fermi energy is at the center of HOMO – LUMO gap. The molecular device configuration considered consists of two vertical gold leads in contact with the horizontal molecules forming ideal bonds with the boron atoms indicated in **Fig.1**

## 3. Results

*3.1 Pure clusters*

**A. Structural properties**. The analysis of the smaller cluster, henceforth named **B6**, was performed by using both basis sets mentioned above in order to make a consistent comparison with $B_{228}H_{24}$, henceforth named **B24**, analyzed only with the minimal basis set. We point out that the results obtained in the two cases are qualitatively equivalent, for this reason, unless specified, the data shown below are referred to the minimal basis set. As far as the boron - boron bond length is concerned, a shorter value has been found in the STO-3G optimized structure, in particular, taking into account cluster **B6**, the average value of this parameter is calculated to be respectively 1.649 Å and 1.638 Å for 6-311G and STO-3G basis set, while a bond length average of 1.629 Å is obtained for the cluster **B24**. The decrease of the bond length average by increasing the size of the cluster is also seen in the Coronene 19, i.e. a molecule of Coronene surrounded by a series of benzene rings, where, by using the same level of calculation, a 0.021 Å decrease of the same parameter is found with respect to the Coronene. It is also interesting to note that the reduction of the bond length takes place in particular in the inner bonds which tend to have the same value.

The cohesive energies, evaluated in the minimal basis set for both clusters, were of 6.437 eV for **B6** and 6.449 eV for **B24**. These values were calculated from:

$$E_{Coh} = E_{Binding}/n \qquad (1)$$

where

$$E_{Binding} = E_{cluster} - \sum E_{ALL\ ATOMS} - \sum E_{ALL\ B-H\ BONDS} \qquad (2)$$

In the expressions above *n* is the number of boron atoms and the value of the B-H bond energy is calculated in the first approximation as 1/3 of the binding energy value of $BH_3$. Structural parameters evaluated are competitive, in terms of stability, with the more stable flat 2D structures considered up to date [32-33].

It is well known that Boron has a variety of compounds containing multicenter bonds, in particular the three-center, two-electron (3c,2e) bond is present in molecules such diborane [34], boron clusters [32] and boron sheets [35]. Previous works have shown that (3c,2e) bonds preclude the formation of boron rings in boron clusters [6,36], whereas, on the other hand, more recently the three-center bonding has been proposed to explain the stability of boron fullerenes [34,37]. This peculiar feature is also seen for **B6** and **B24,** while it is not present in Coronene and its larger clusters such Coronene 19, Coronene 37 and Coronene 61. Hence it is logical to assume that, as for boron fullerenes, it plays a pivotal role in maintaining a 2D stable planar structure. The presence of (3c,2e) bonds can be evaluated by means of the Mayer Bond Order indices, calculated from the canonical MOs in the canonical AO basis [34,38-40], which, for closed-shell species with 3 center bonds involving the atoms A, B and C, can be expressed as follows:

$$I_{ABC} = \sum_{\alpha}^{A} \sum_{\beta}^{B} \sum_{\gamma}^{C} (PS)_{\alpha\beta} (PS)_{\beta\gamma} (PS)_{\gamma\alpha} \qquad (3)$$

where P is the total density matrix and S is the overlap matrix. The bond order indices of three-center bonds are positive with a theoretical maximum of $\approx 0.296$.

With reference to **Fig 2**, we report in **Tab. 1** the more relevant values of ***I*** for both clusters, in general one can affirm that each boron is involved at least in two different three-center bonds, i.e. each boron is directly linked at least with four boron atoms.

**B. Electronic properties.** It has been suggested that the anomalous stability of the boron planar clusters depends on the aromaticity which arises from the delocalization of π-electrons and involves unoccupied $2p_z$ Atomic Orbitals [11-12,14]. As it will be discussed below, **B6** and **B24** show these features; before analyzing in detail we underline that a more extended electronic delocalization gives rise to a smaller GAP in carbon clusters [41], described as the HOMO – LUMO energy difference. In accordance with these calculations, the GAP values obtained for **B6** and **B24** are 1.33 eV and 1.17 eV respectively; their density of states (DOS) as a function of energy (eV) is shown at **Fig. 3**. At energy E the density of states is written as:

$$DOS(E) = \sum_i \delta(E - \varepsilon_i)$$

where the summation index *i* goes over all energy levels and δ is the Delta function.

From **Fig. 3** we note that: (a) both curves show a similar profile, cluster **B24** has a larger density of states, while differently from cluster **B6**, it shows a zero value of the DOS only within the HOMO-LUMO gap; (b), the composition of the molecular orbitals, calculated by means of Mulliken Population Analysis, reported in **Tab 2**, clearly highlights how this contributes to the HOMO of both clusters, shown in the insets of **Fig 3**. and to their nearer molecular orbitals, are mainly due to the $p_z$ atomic orbitals, confirming the stabilizing effect of π-delocalization. This result is in agreement with the one evidenced in Coronene, whose HOMO and DOS are reported, in **Fig.3**; (c), as already observed for the set of Carbon clusters previously studied [41], the peaks near the HOMO energy can be joined by an almost straight line, reproducing the linear dependence shown by the infinite system near the Fermi level.

Now we turn to aromaticity which, as already mentioned, is considered as the basis of stability for boron planar clusters. Szwacki et al. [15] have discussed about the regions of aromaticity of borozene and since this molecule can be indicated as the embryo of **B6** and **B24**, we find necessary to investigate this aspect.

**Fig. 4** shows the plot of the nucleus independent chemical shift (NICS), which represents the magnetic criterion to evaluate the ring current for cluster **B24**. Negative value of NICS arise when diatropic ring current dominates, meaning that the system considered is aromatic, on the other hand a paratropic current gives rise to a positive value of NICS, therefore the corresponding system is antiaromatic. From **Fig. 4** it is evident that inner bonds give rise to a paratropic current inside the round areas which can be considered as the expansion of the inner triangle antiaromatic area found in borozene [15] , whereas a flow of diatropic current is homogeneously present in the rest of the cluster.

In **Fig. 5** the low bias differential conductance of the two clusters ideally contacted with two gold leads is shown. The symmetry of the plots reflects the assumed symmetry in the device configuration. A semiconductor-like behavior is evidenced in both structures. However, the zero bias differential conductance is one order of magnitude higher for the B24 cluster, this is due not only to the lower HOMO – LUMO gap but also to the larger value of the DOS. Delocalized (along the cluster) $p_z$ molecular orbitals allow an efficient charge transport through the cluster for larger bias (of the order of the gap) and a diode-like characteristic can be observed.

*3.2 Clusters with defects.*

When impurities are taken into account, regardless of both the size of the clusters and the type of defects considered, differently from carbon clusters, the planar structure is compromised. It is worth noticing that **B6** and **B24** show a high level of connectivity, in fact, in addiction to the three center – two electron bond, a relevant value of the two center Wiberg Bond Order index [42] has been calculated, involving various atomic pairs; with reference to **Fig. 2**, a value of 0.389 is found for pairs 9-13, 17-18 and 25-28.

The removal of a boron atom gives rise to a heavy weakening of the connectivity and, as a consequence, planar geometry is not maintained; similarly, since that carbon and nitrogen form shorter bonds than boron, the same effect is seen when substitutional atoms are introduced.

The discussion above indicates that in on one hand the pure 2D clusters can be considered stable, while on the other hand the presence of impurities brings to a large deformation; this aspect might be one of the main difficulties in synthesizing large sized 2D clusters precluding their potential applications.

**4. Conclusions**

In this work first principles and semiempirical calculations were carried out to investigate both structural and electronic properties of two clusters obtained by condensation of 6 and 24 borozene molecules, considered as the analogues of Coronene and Coronene 19. Structural effects induced by the introduction of impurities such as a) vacancies and b) substitutional atoms, are also taken into account.

Calculations predict a planar geometry for the pure clusters. Both (3c-2e) bonds and wide regions of aromaticity contribute to this stabilization, with cohesive energy values that are comparable with the most stable boron clusters considered to date. Due to the high connectivity among boron atoms, the planar geometry is compromised when impurities are introduced; this feature strongly differentiates **B6** and **B24** from carbon compounds such Coronene [40] .

Density of states spectra evidence a small gap, which decreases by increasing the cluster size, suggesting, at variance with carbon clusters, a semiconducting character in small sized clusters; furthermore the population analysis shows that the main contribution to the molecular orbitals near the GAP is due to π-bonds which derive from $p_z$ orbitals.

Calculated low-bias differential conductance for these structures confirms this semiconductor like character.

**Acknowledgements**

G. Forte wishes to thank the Consorzio Interuniversitario Cineca for the computational support.

**Tab.1** Principal 3 center - 2 electron Mayer bond order indices for clusters B6 and B24. Atom labels are referred to **Fig.2**

| Atoms | B6 | B24 |
|---|---|---|
| 1-2-4 | 0.238 | 0.243 |
| 1-3-4 | 0.192 | 0.185 |
| 2-4-5 | 0.184 | 0.192 |
| 3-4-7 | 0.184 | 0.192 |
| 3-6-7 | 0.185 | 0.181 |
| 4-5-8 | 0.184 | 0.192 |
| 5-8-9 | 0.185 | 0.189 |
| 7-8-12 | 0.218 | 0.218 |
| 7-11-12 | 0.182 | 0.180 |
| 8-9-13 | 0.129 | 0.121 |
| 8-12-13 | 0.182 | 0.178 |
| 9-13-15 | 0.129 | 0.115 |
| 9-15-16 | 0.182 | 0.191 |
| 13-14-15 | 0.182 | 0.178 |
| 19-20-23 |  | 0.212 |
| 20-21-24 |  | 0.213 |
| 23-26-27 |  | 0.184 |
| 23-24-27 |  | 0.217 |
| 24-27-28 |  | 0.184 |

**Tab.2** Energies and percentual contributes of $p_z$ orbitals to the composition of HOMO, LUMO and their nearest MOs.

|  | B6 | | B24 | |
|---|---|---|---|---|
|  | E (eV) | % $p_z$ | E (eV) | % $p_z$ |
| HOMO-3 | -4.73 | 0.36 | -3.67 | 100.00 |
| HOMO-2 | -4.52 | 99.87 | -3.67 | 99.94 |
| HOMO-1 | -4.52 | 99.80 | -3.43 | 99.99 |
| HOMO | -3.56 | 99.97 | -3.28 | 99.99 |
| LUMO | -2.23 | 99.81 | -2.12 | 100.00 |
| LUMO +1 | -1.50 | 91.19 | -2.06 | 100.00 |
| LUMO +2 | -1.36 | 95.71 | -1.89 | 99.93 |
| LUMO +3 | -1.36 | 99.94 | -1.89 | 99.94 |

**Fig.1** Clusters $B_{60}H_{12}$ (left) and $B_{228}H_{24}$ (right) obtained after geometry optimization. Highlighted in red are the molecule of borozene (left) and the cluster $B_{60}H_{12}$ (right). In yellow are highlighted the boron atoms in contact with the gold leads in the transport calculations

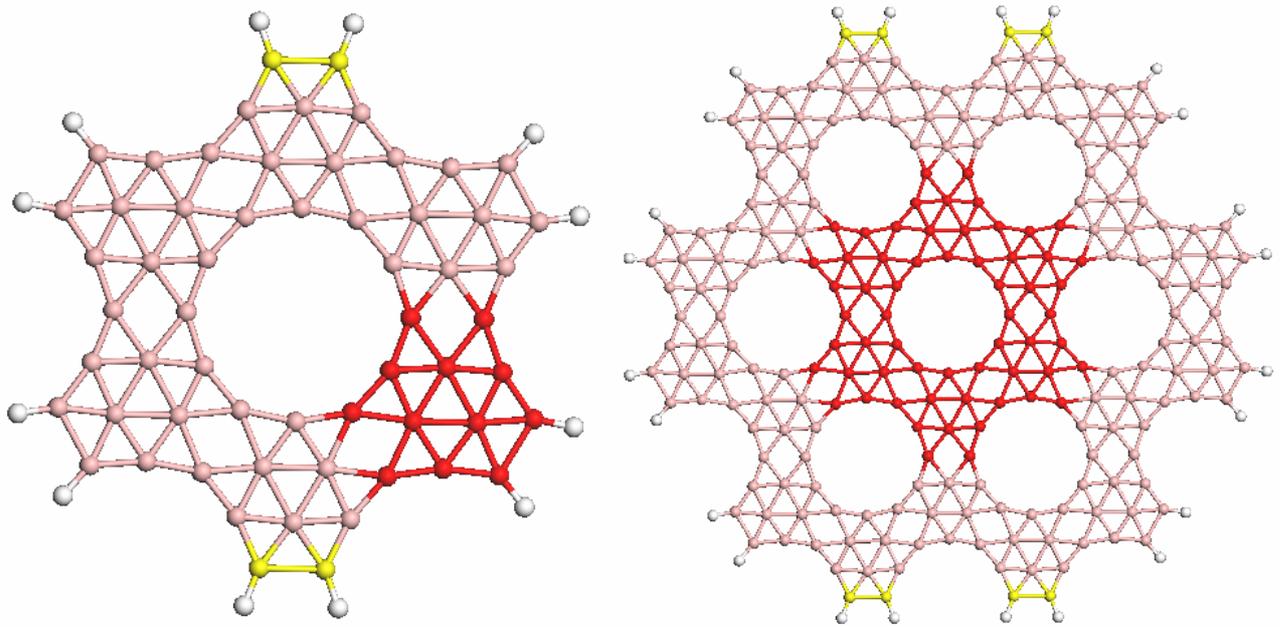

**Fig.2** A section of the cluster $B_{228}H_{24}$; labels from 1 to 17 are also referred to $B_{60}H_{12}$

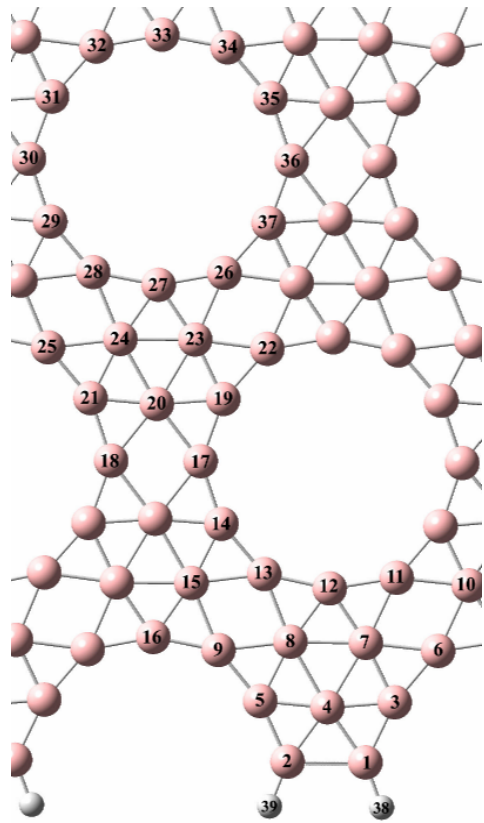

**Fig.3** Above: Density of states and HOMO (in the insets) for cluster B6 (black line) and B24 (red line). Below: Density of states and HOMO (in the inset) for Coronene

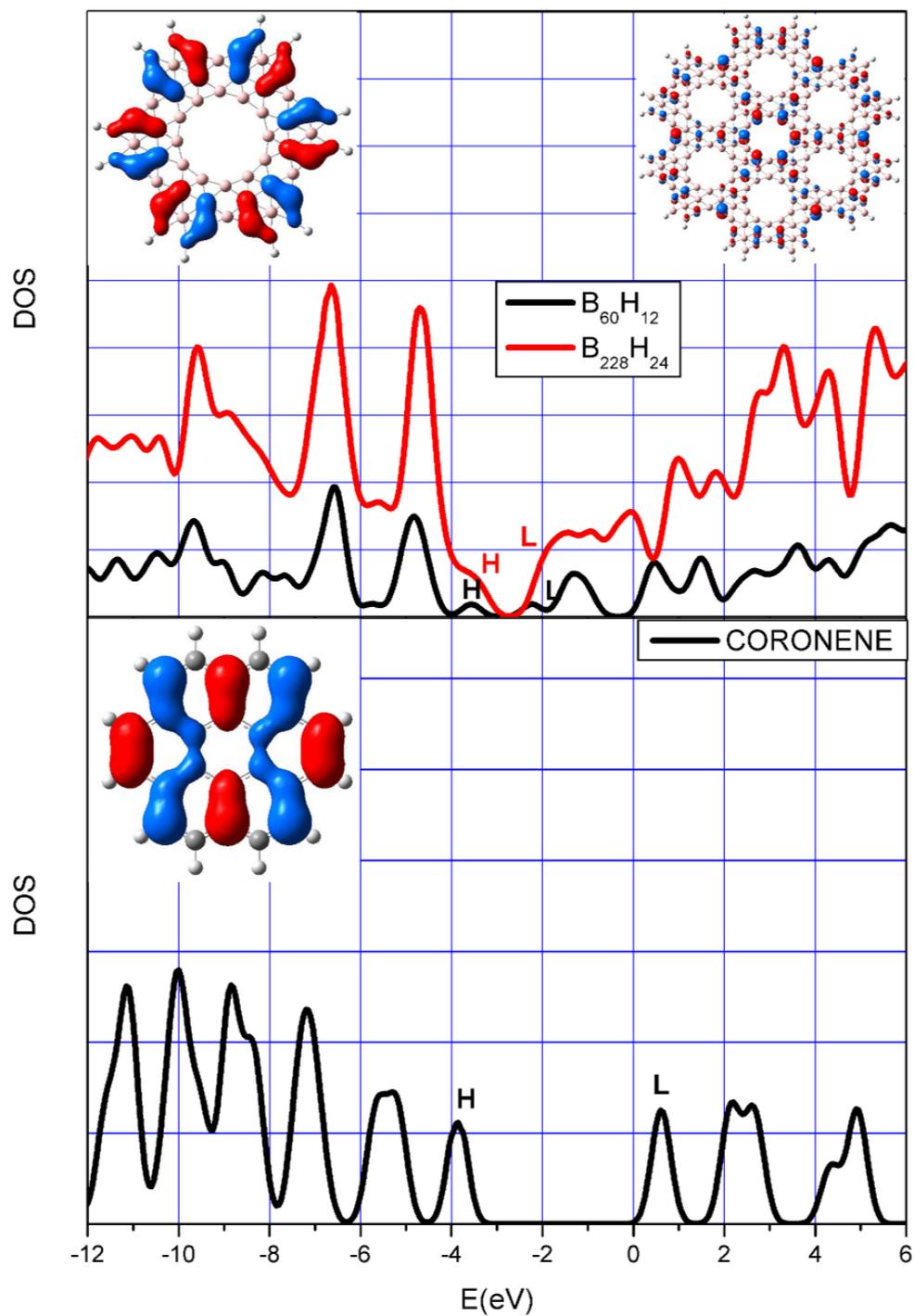

**Fig.4** In plane contour plot of NICS (X,Y) for B24. The step size of the ghost atoms is about 1 Å.

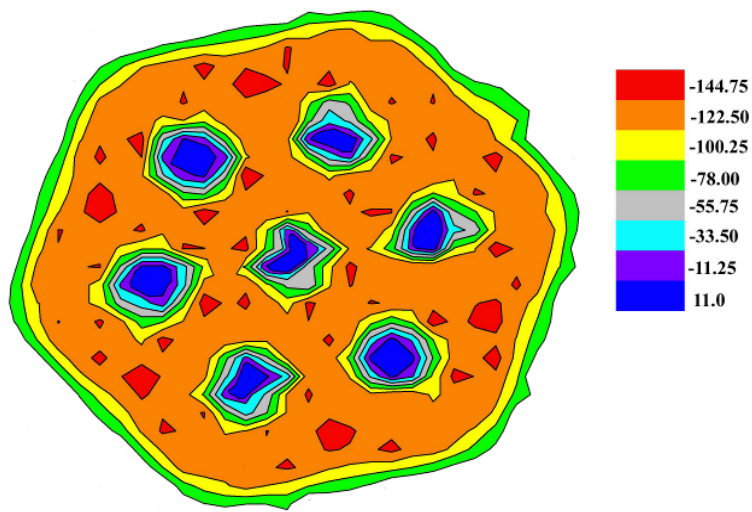

**Fig.5** Caption red line (black line): low bias differential conductance of the B6 (B24) molecule ideally contacted with gold leads in the two-terminal configuration.

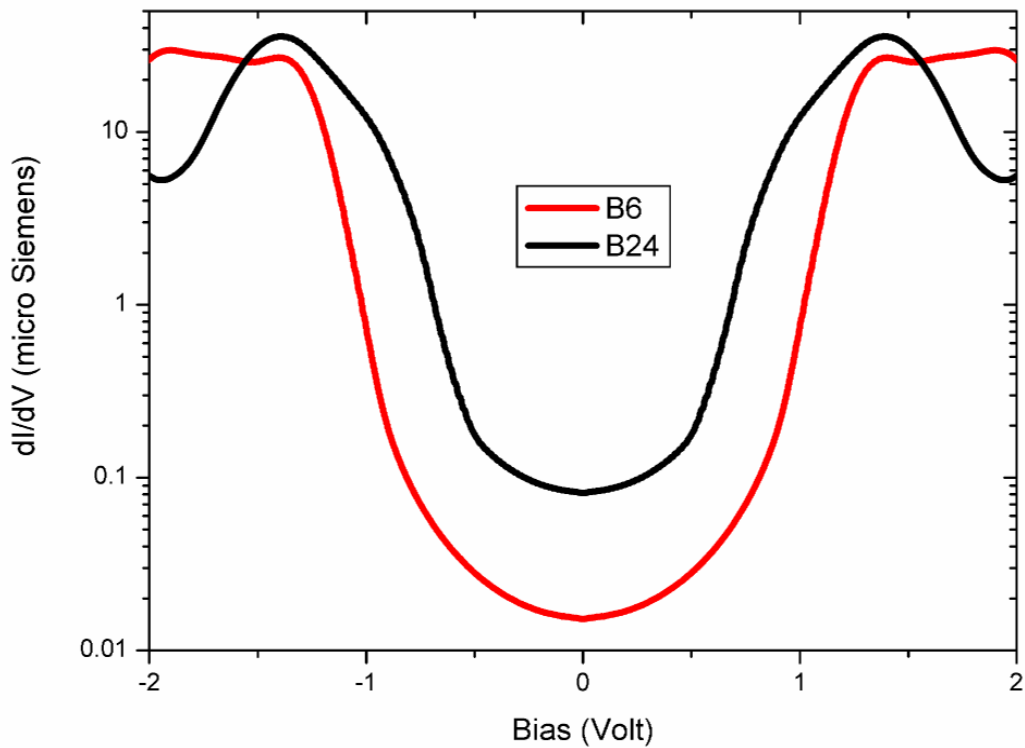